\documentclass{article}

\usepackage{arxiv}
\usepackage[utf8]{inputenc} 
\usepackage[T1]{fontenc}    
\usepackage{hyperref}       
\usepackage{url}            
\usepackage{booktabs}       
\usepackage{amsfonts}       
\usepackage{nicefrac}       
\usepackage{microtype}      
\usepackage{lipsum}		
\usepackage{graphicx}
\usepackage{doi}
\usepackage{amsmath}
\newenvironment{assumption}{\textbf{Assumption}}{\par}

\title{A Comprehensive Framework for Estimating Aircraft Fuel Consumption Based on Flight Trajectories}

\author{ \href{https://orcid.org/0000-0003-2170-830X}{\hspace{1mm}Linfeng ~Zhang} \\
	School of Logistics\\
	Beijing Wuzi University\\
	Beijing, China \\
	\texttt{zlflbl@163.com} \\
	\And
	\href{https://orcid.org/0000-0001-9574-7673}{\hspace{1mm}Alex L.~Bian} \\
	TravelSky Mobile Technology Limited\\
	Beijing, China \\
	\texttt{bianlei@pku.edu.cn} \\
        \And
	\href{https://orcid.org/0000-0002-0775-5102}{\hspace{1mm}Changmin~Jiang} \\
	Department of Logistics and Maritime Studies\\
	The Hong Kong Polytechnic University\\
	Hong Kong, China \\
	\texttt{changmin.jiang@polyu.edu.hk} \\
       \And
	\href{https://orcid.org/0000-0002-5123-7556}{\hspace{1mm}Lingxiao~Wu}\thanks{Corresponding Author.} \\
	Department of Aeronautical and Aviation Engineering\\
	The Hong Kong Polytechnic University\\
	Hong Kong, China \\
	\texttt{maillingxiao-leo.wu@polyu.edu.hku} \\
}

\renewcommand{\shorttitle}{Aircraft Fuel Consumption}

\begin{document}
\maketitle

\begin{abstract}
Accurate calculation of aircraft fuel consumption plays an irreplaceable role in flight operations, optimization, and pollutant accounting. Calculating aircraft fuel consumption accurately is tricky because it changes based on different flying conditions and physical factors. Utilizing flight surveillance data, this study developed a comprehensive mathematical framework and established a link between flight dynamics and fuel consumption, providing a set of high-precision, high-resolution fuel calculation methods. It also allows other practitioners to select data sources according to specific needs through this framework. The methodology begins by addressing the functional aspects of interval fuel consumption. We apply spectral transformation techniques to mine Automatic Dependent Surveillance-Broadcast (ADS-B) data, identifying key aspects of the flight profile and establishing their theoretical relationships with fuel consumption. Subsequently, a deep neural network with tunable parameters is used to fit this multivariate function, facilitating high-precision calculations of interval fuel consumption. Furthermore, a second-order smooth monotonic interpolation method was constructed along with a novel estimation method for instantaneous fuel consumption. Numerical results have validated the effectiveness of the model. Using ADS-B and Aircraft Communications Addressing and Reporting System (ACARS) data from 2023 for testing, the average error of interval fuel consumption can be reduced to as low as $3.31\%$, and the error in the integral sense of instantaneous fuel consumption is $8.86\%$. These results establish this model as the state of the art, achieving the lowest estimation errors in aircraft fuel consumption calculations to date.
\end{abstract}


\section{Introduction}\label{sec:Intro}
Computation of aviation fuel consumption stands as a pivotal element in the realms of aerospace engineering and environmental science, holding profound significance for both government regulators and airline operators. From the perspective of airline operations, fuel costs represent a significant portion of operating expenses for airlines~\cite{khan2021prediction}. The uncertainty of fuel consumption both increase the operating costs of airlines. Due to improper loading, more than two hundred million in fuel is wasted annually~\cite{kang2018evaluating}. That means precise fuel consumption calculations help in formulating more effective fuel usage policies, advancing sustainable aviation technologies, and improving overall efficiency in the aerospace sector. From the perspective of environmental protection and government regulators, understanding fuel consumption patterns is essential for developing strategies to mitigate environmental impact. Unlike ground transport, which can reduce emissions and carbon footprint through electrification and route optimization \cite{fernandez2022arc,liu2023branch}, the aviation industry faces significant challenges in adopting low-emission technologies due to the substantial quantities of energy-dense fuel necessary for long-distance travel~\cite{speizer2024integrated}. Although the existing methods for reducing aviation emissions involve using affordable Sustainable Aviation Fuels~\cite{jiang2023recent}, or improving aircraft fuel efficiency~\cite{brueckner2024airlines}, they inherently assume a reduction in fuel consumption. This suggests that precise fuel measurement is crucial for environmental conservation and enables government regulators to make more accurate estimations of emissions reductions and to assess the effectiveness of various emission control measures more effectively.

In aviation, fuel consumption is typically categorized into two types: interval fuel consumption and instantaneous fuel consumption. Interval fuel consumption refers to the amount of fuel consumed over a specific period of time or distance. And instantaneous fuel consumption measures the rate at which fuel is being used at any given moment. It is typically expressed in terms of fuel flow (like gallons or liters per hour) and provides a snapshot of how much fuel an aircraft is using under specific operating conditions, such as different flight attitude. Interval fuel consumption, as a key indicator for assessing fuel efficiency and economic performance of flights, is crucial for airline route planning, aircraft selection, and economic evaluation~\cite{wells2023reformulating}. By optimizing interval fuel consumption, airlines can not only reduce costs and enhance competitiveness but also diminish environmental impacts. Moreover, calculating instantaneous fuel consumption allows for tracking carbon emissions during each flight phase, providing higher spatial and temporal resolution, beneficial for grid-based accounting of pollutant emissions, thus aiding in more accurately assessing the environmental impact of flights and informing strategies to reduce carbon emissions~\cite{jiang2023recent}.

The importance of calculating aviation fuel consumption is evident as it directly impacts the economic efficiency and environmental footprint of airlines~\cite{timmis2015environmental,zhu2018study,seymour2020fuel}. However it faces several challenges: Firstly, there is a lack of precision. It is difficult for theoretical models to account for actual aircraft operational conditions when calculating fuel consumption.
Secondly, data acquisition presents significant challenges. These challenges include obtaining precise data and managing large volumes of information. Devices like the Aircraft Communications Addressing and Reporting System (ACARS) , Quick Access Recorder (QAR) and Flight Data Recorder (FDR) do capture extensive operational data, such as instantaneous fuel consumption. However, access to this data is only available after the aircraft has landed~\cite{chang2015examination}, and it also faces challenges such as irregular data uploads and a limited number of data points. Moreover, such data is often considered a trade secret and should be handled with confidentiality, not disclosed indiscriminately to the public.
Lastly, there is a lack of standardization in fuel calculations, which compromises the generalizability of most algorithms. Fuel consumption is influenced by numerous variables, including flight altitude, time, speed, environmental conditions, and air traffic flow. Moreover, even identical flights may operate at different altitudes on different days due to random variations, further complicating consistent fuel calculation~\cite{pagoni2017calculation,brueckner2024airlines}. 

To address these challenges, this study introduces a comprehensive and widely applicable mathematical theory, along with a data framework for aircraft fuel consumption calculation. In our practice, we utilized Automatic Dependent Surveillance Broadcast (ADS-B) and ACARS data to achieve precise modeling of the relationship between flight attitude and fuel consumption. ADS-B is low-cost data which can be used to identify aircraft behaviors such as acceleration, deceleration, climb, and descent. ACARS data, being real-time air-to-ground link data, ensures that the information is updated in real time during flight. The framework of this study is highly versatile, allowing researchers and practitioners to select appropriate data sources based on specific needs. The findings of this research are not only theoretically innovative but also demonstrate significant practical value and potential in real-world applications. Overall, the contributions of this study are as follows.

\begin{enumerate}
 \item This study provides high-precision, high-resolution fuel calculations based on ADS-B data mining of aircraft dynamics. The use of ADS-B data allows for a granular examination of flight dynamics, which providing a rich dataset to analyze minute-by-minute changes in flight conditions and their effects on fuel consumption. The average error of interval fuel consumption can be reduced to as low as 3.31$\%$, and the error in the integral sense of instantaneous fuel consumption is 8.86$\%$. 
 \item This study achieves rapid and accurate calculations, along with more usable fuel data. We have developed a model fuel calculation database for 11 commonly used aircraft types, which covers 90.06$\%$ of the flights in China. This enables researchers to independently calculate fuel consumption for different flight distances using publicly available ADS-B and aircraft model data. For other aircraft types, researchers or engineers can use our models to create their own database. 
 \item This research introduces a broadly applicable framework and a rigorous mathematical theory system that systematically verifies the relationship between the trajectory of transportation vehicles and their energy consumption for the first time. Due to the universality of this theoretical framework, it is applicable not only to commercial airliners but also can be extended to other types of vehicles, such as space consumption and ships' fuel usage. This study not only fills a gap in existing research, providing valuable theoretical support and practical guidance for the field, but also offers a new methodological approach to optimizing energy efficiency and assessing environmental impacts of transportation vehicles. 
\end{enumerate}

The rest of this paper is organized as follows. Section \ref{sec:review} reviews the most relevant literature on  estimating the fuel consumption. Section \ref{sec:formulation} presents the framework for computing both interval and instantaneous fuel consumption of aircraft. Error analysis is provided in Section \ref{sec:erroranalysis}. We show an application in Section \ref{sec:application}. Then the conclusions are following in Section \ref{sec:Conclusion}.

\section{Literature Review}\label{sec:review}

There are multiple methods to estimate the fuel consumed by an aircraft. A summary of common calculation methods for various models is provided in \cite{kuhn2023fuel}. Combining the content from this book and other application articles, we categorize the methods of calculating fuel consumption into three groups: mathematical models based on the performance and physics, data driven method supported by newer techniques, such as machine learning or the usage of big-data, and hybird method.

\subsection{Mathematical models}

Mathematical models such as Specific Air Range (SAR),  EEA Master Emission Calculator,  Payload-Range Diagram,  Base of Aircraft Data (BADA),  Handbook Methods, energy-balance (EB) are based on complex mathematical formulation and provide a better physical understanding of the system~\cite{khan2021prediction,nuic2010bada}.
Many scholars conduct research on aviation fuel consumption or fuel efficiency calculations based on those methods, for excample improved SAR methods ~\cite{babikian2002historical,yutko2011approaches,lovegren2011estimation,jensen2013commercial}, or advanced fuel burn model (AFBM)~\cite{khan2021prediction}.

These methods highlight the ongoing challenges in aviation fuel consumption calculations:
1.the unavailability of specific aircraft parameters and data to the public~\cite{phase1997enhancements,yanto2018aircraft, zhang2020bayesian}. 2.the definition of flight phases—including takeoff, climb, descent, and landing—poses difficulties due to their variable nature, which lead to those method more accurate for the cruise phase, whereas for terminal phases, it losses accuracy~\cite{pagoni2017calculation,turgut2017investigating}. 3.the accuracy of these calculations is often limited, impacting the reliability of fuel consumption estimates~\cite{zhang2020bayesian,gongzhang2022improving}.

\subsection{Data-driven method}

With the advent of the data era, there has been a growing interest in data-driven research on aviation fuel consumption to address the limitations in the parameterization of theoretical models. Metric Value is provides a concise fuel efficiency assessment for easy comparison between different aircraft and flights ~\cite{kuhn2023fuel}.
With the adoption of Quick QAR and Flight Data FDR technologies ~\cite{todd2002flight, chati2018data,wang2014analysis,huang2022estimation}, historical logs capturing parameters such as vertical speed, gross weight, and fuel flow are increasingly utilized in data-driven approaches to estimate fuel consumption. In parallel, \cite{kang2021quantile} advocate for the use of quantile regression-based machine learning methods to mitigate uncertainties in aircraft fuel burn, enhancing the accuracy of fuel consumption estimates. Moreover, neural network topology and hybrid machine learning models have developed to predict aircraft fuel consumption~\cite{de2013machine,zhu2021flight,metlek2023new,lin2024fuel}.

This existing related works based on data driven method achieve the fuel consumption task based on the aircraft performance and post-flight trajectory, serving as the post-analysis tool for airline operations. However, the variability inherent in data-driven models, due to the dynamic nature of physical rules across different spatial and temporal conditions, often hampers their ability to effectively capture relationships within training data, resulting in poor generalization capabilities to new, unseen datasets ~\cite{uzun2021physics}. The difficulty in data acquisition further complicates the effectiveness of these models. For instance, accessing comprehensive FDR data is challenging due to safety and privacy concerns, and often, the available data may be outdated or lack timeliness~\cite{korba2023optimizing}. Additionally, the methods used to segment flight data, such as during the Landing Takeoff Cycle (LTO), can be too coarse, making it difficult to accurately identify and compute specific flight attitudes like cruising and circling maneuvers. These challenges highlight the need for more refined data handling and analysis techniques in the modeling of aviation dynamics~\cite{mumaw2020analysis}.

\subsection{Hybrid method}

The hybrid approach blends mathematical and data-driven models. The Bathtub Curve, referenced in ~\cite{burzlaff2017aircraft,ekici2023performance}, calculates fuel burn for specific voyages and generates graphs using Excel. \cite{phase1997enhancements} developed an aircraft fuel consumption model (AFCM) using EB and neural networks to simulate point fuel burn, while ~\cite{rodriguez20194d} introduced a methodology employing BADA and Monte Carlo simulations. \cite{lyu2020flight} utilized a hybrid data-driven and physics-based model to map flight operational data to aircraft performance, estimating fuel consumption and monitoring flight conditions. Previous studies by ~\cite{zhang2020bayesian} and ~\cite{pagoni2017calculation} categorized trajectory prediction literature into two broad categories: mathematical and data-driven models.
\cite{baumann2021evaluation} analyzed fuel efficiency in Airbus A320 using simulated data and machine learning method, examining the impact of improved winglets. \cite{gongzhang2022improving} introduced an ensemble learning-based method that significantly boosts aircraft fuel consumption prediction accuracy.

The hybrid model merges theoretical and data-driven approaches, enhancing fuel consumption calculation accuracy but faces limitations due to the need for high-quality data and constraints from theoretical assumptions.

In summary, current inaccuracies in aviation fuel predictions stem from inadequate analysis of flight trajectory data and the absence of a robust methodology for its effective use. Moreover, the lack of precise methods for interval fuel calculation and ineffective smoothing techniques hinders accurate instantaneous fuel consumption computation.

\section{Mathematical Formulation to Calculting Aircraft Fuel Consumption}\label{sec:formulation}

In this section, a high-precision, high-resolution fuel calculations method is provided based on ADS-B data mining of aircraft dynamics, deep learning, and second-order smooth monotonic interpolation. We primarily utilize a large volume of low-cost ADS-B data and relatively sparse high-cost ACARS data. ADS-B data includes messages such as Airborne Position Message and Airborne Velocity Message, providing detailed information on aircraft flight parameters. ACARS data records partial interval fuel quantity changes. However, ACARS data uploads are irregular, and the number of data points is limited.

We propose a mathematical framework to calculate both interval and instantaneous fuel consumption of aircraft. The framework consists of three main steps:
\begin{enumerate}
    \item Using ADS-B data to determine the flight profile, which involves identifying aircraft behaviors such as acceleration, deceleration, climb, and descent, and establishing their theoretical relationships and parameters with fuel consumption.
    \item Fitting coefficients of the relationship between flight profiles and fuel consumption using available interval fuel consumption data.
    \item Calculating instantaneous fuel consumption through monotonic and smooth interpolation.
\end{enumerate}

\subsection{Basic assumptions and Mathematical Approach}

Based on our previous discussions and a review of the literature, several key principles regarding the working mechanism of aircraft engines have been established. For a specific aircraft, the engine's fuel consumption primarily serves the following purposes: {Maintaining Altitude, Climbing or Descending, Maintaining Airspeed, Accelerating or Decelerating.

The instantaneous fuel consumption of an aircraft is not only related to inherent physical parameters such as aircraft type and age, but also directly correlates with the aircraft's altitude, velocity, altitude change rate, and velocity change rate.

\begin{equation}
	\xi _{ij}(t)=P(x_{t}=i,x_{t+1}=j|y,v,w;\theta)= {\frac {\alpha _{i}(t)a^{w_t}_{ij}\beta _{j}(t+1)b^{v_{t+1}}_{j}(y_{t+1})}{\sum _{i=1}^{N} \sum _{j=1}^{N} \alpha _{i}(t)a^{w_t}_{ij}\beta _{j}(t+1)b^{v_{t+1}}_{j}(y_{t+1})}}
\end{equation}

Let the instantaneous fuel consumption of an aircraft with $d$ fixed parameters $\Xi=\{\xi_1,\xi_2,\dots,\xi_d\}$ be denoted as $q(t;\Xi) = q(t;\xi_1,\xi_2,\dots,\xi_d)$, where the parameters $\xi_j$ can refer to intrinsic aircraft properties such as model, wingspan, age, empty weight, and maximum takeoff weight, the instantaneous fuel consumption of the aircraft can be written as
\begin{equation}
    \label{equ:model:ranyouq}
    q(t) = q(h(t),h'(t),h''(t),v(t),v'(t),v''(t);\Xi),
\end{equation}
where Eq.(\ref{equ:model:ranyouq}) does not explicitly contain time $t$, reads $\displaystyle{\frac{\partial q}{\partial t}  = 0}$. Here, $h(t)$ and $v(t)$ are the altitude and speed of the aircraft as functions of time, respectively. The first and second derivatives $h'(t)$, $v'(t)$ and $h''(t)$, $v''(t)$ represent the rates of change of altitude and speed.

Integrating Eq.(\ref{equ:model:ranyouq}) over $[0,t_0]$, we obtain the interval fuel consumption, denoted
\begin{equation}
    \label{equ:model:qujianQ1}
    \begin{aligned}
        Q & =  Q(h(t),v(t);t_0,\Xi)\\
          & =  \int_0^{t_0}q(h(t),h'(t),h''(t),v(t),v'(t),v''(t);\Xi)dt.
    \end{aligned}
\end{equation}
$Q$ is a functional of the functions $h(t)$ and $v(t)$, parameterized by $t_0$ and $\xi_1,\xi_2,\dots,\xi_d$. Given a set of $h(t)$ and $v(t)$, there exists a corresponding interval fuel consumption $Q$.

\begin{assumption}[Continuity]
\label{HYP:Continuity}
For a specific aircraft with $\Xi$, if the altitudes $h(t)$ and velocities $v(t)$ are identical in two flights, meaning that the flight profiles are exactly the same, then the fuel consumption for both flights will be exactly the same. If there are slight differences between the altitude and velocity profiles of two different flights, for example, if the altitude and velocity of the first flight are $h(t)$ and $v(t)$, and for the second flight they are $h(t)+\delta h(t)$ and $v(t)+\delta v(t)$, then
\begin{equation}
    \label{equ:model:qujianQ1lianxu1}
    \lim_{\delta h(t)\rightarrow 0}\lim_{\delta v(t)\rightarrow 0} \delta Q  = 0.
\end{equation}
Here $\delta Q$ represents the variation of the functional $Q$, denoted as
\begin{equation}
    \label{equ:model:qujianQ1lianxu2}
    \delta Q  =  Q(h(t)+\delta h(t),v(t)+\delta v(t);t_0,\Xi)-
         Q(h(t),v(t);t_0,\Xi)
\end{equation}
Eq.(\ref{equ:model:qujianQ1lianxu2}) indicates that the functional $Q$ is continuous with respect to $h(t)$ and $v(t)$.
\end{assumption}

In fact, we often lack effective approaches when dealing with functional problems. One way to simplify these problems is through function approximation. Suppose we have a set of basis functions $\{ r_j(t)| t\in [0,t_0], j = 0,1,\dots \}$ in the function space over the interval $[0,t_0]$, we consider a good approximation of the functions $h(t)$ and $v(t)$ over $[0,t_0]$.
\begin{equation}
    \label{equ:model:htbijin}
    h(t) = \sum_{j=0}^{N_h}\alpha_jr_j(t)+\epsilon_h(t,N_h), \quad v(t) = \sum_{j=0}^{N_v}\beta_jr_j(t)+\epsilon_v(t,N_v).
\end{equation}
Here, $N_h$ and $N_v$ are truncation radii. If Eq.(\ref{equ:model:htbijin}) provide a good approximation of the functions $h(t)$ and $v(t)$, we have $\displaystyle{\lim_{N_h\rightarrow \infty}\epsilon_h(t,N_h)=0}$ and $\displaystyle{\lim_{N_h\rightarrow \infty} \epsilon_v(t,N_v) = 0}$. $\alpha_j$ and $\beta_j$ are the coefficients obtained from the approximation process, and $r_j(t)$ are the basis functions used for approximating $h(t)$ and $v(t)$, respectively. If these approximations are accurate, then the functional $Q$ can then be expressed as
\begin{equation}
    Q = Q\left(\sum_{j=0}^{N_h}\alpha_jr_j(t)+\epsilon_h(t,N_h),\sum_{j=0}^{N_v}\beta_jr_j(t)+\epsilon_v(t,N_v);t_0,\Xi\right)
\end{equation}
When $N_h$ and $N_v$ are sufficiently large, the functions $\epsilon_h(t,N_h)$ and $\epsilon_v(t,N_v)$ become sufficiently small. Considering the Assumption \ref{HYP:Continuity} and taking into account the continuity condition (\ref{equ:model:qujianQ1lianxu2}), we have an approximation for the functional $Q$ as follows:
\begin{equation}
    \label{equ:model:qujianQ2}
    \begin{aligned}
        Q & =  Q(h(t),v(t);t_0,\Xi)\\     &=Q\left(\sum_{j=0}^{N_h}\alpha_jr_j(t)+\epsilon_h(t,N_h),\sum_{j=0}^{N_v}\beta_jr_j(t)+\epsilon_v(t,N_v);t_0,\Xi\right)\\
          & \approx  \int_0^{t_0}q\left(\sum_{j=0}^{N_h}\alpha_jr_j(t),\sum_{j=0}^{N_h}\alpha_j\dot{r}_j,\sum_{j=0}^{N_h}\alpha_j\ddot{r}_j,\sum_{j=0}^{N_v}\beta_jr_j(t),\sum_{j=0}^{N_v}\beta_j\dot{r}_j,\sum_{j=0}^{N_v}\beta_j\ddot{r}_j;t_0,\Xi\right)dt.
    \end{aligned}
\end{equation}
 The derivatives with respect to $t$ ultimately fall onto the basis functions $r_j(t)$. Since $r_j(t)$ are known functions over and the expression for the instantaneous fuel consumption $q$ does not explicitly depend on time $t$, we can perform a formal integration on the right side of Eq.(\ref{equ:model:qujianQ2}) as follows:
\begin{equation}
    \label{equ:model:qujianQ3}
    Q \approx Q(\mathbf{A}_{N_h},\mathbf{B}_{N_v};t_0,\Xi).
\end{equation}
Here, the vectors $\mathbf{A}_{N_h}= (\alpha_1,\alpha_2,\dots,\alpha_{N_h})$, and $\mathbf{B}_{N_v} = (\beta_1,\beta_2,\dots,\beta_{N_v})$, respectively.

Thus, the interval fuel consumption functional $Q$ has been simplified into a multi-variable real-valued function involving $d+1$ parameters $(t_0,\xi_1,\xi_2,\dots,\xi_d)$, with respect to $(\alpha_1,\alpha_2,\dots,\alpha_{N_h},\beta_1,\beta_2,\dots,\beta_{N_v})$. Consequently, the calculation of interval fuel consumption is transformed into a fitting problem for a multi-variable real-valued function with parameters. There are numerous deep learning techniques available for addressing the fitting problem of such multi-variable real-valued functions with parameters, including DNN (Deep Neural Networks), Wide and Deep Methods, DeepFM, and others.
If we can determine the interval fuel consumption for any given segment of an ADS-B sequence using the above results, we can then calculate the instantaneous fuel consumption. Here, we assume that the interval fuel consumption for any two segments can be computed. Thus, we have the following sequence of $N$ interval fuel consumption:
\begin{equation*}
    (t_0,Q_0),(t_1,Q_1),\dots,(t_N,Q_N).
\end{equation*}
For a given sequence with specific parameters $\Xi$, considering the positivity of fuel consumption $Q_j<Q_k, \forall 0\leq j<k\leq N$, we construct a monotonically increasing function $Q(t;\Xi)$ defined over $[0,t_N]$ such that $Q(t_j;\Xi) = Q_j$. If $Q(t)$  is twice continuously differentiable over $(0,t_N)$, the instantaneous fuel consumption reads $\displaystyle{q(t;\Xi) = \frac{d}{dt}Q(t;\Xi)}$. This ensures $q(t;\Xi)$ is well-defined and can be accurately calculated at any point within the interval.

Thus, we can now accurately calculate the interval fuel consumption and instantaneous fuel consumption based on the flight's ADS-B data. The remaining sections of this chapter will provide a detailed explanation of the approximation techniques used for altitude and velocity sequences, the fitting techniques for multi-variable real-valued functions, and the methods for smooth function interpolation that preserve monotonicity.

\subsection{Time Sequences Approximation}

Almost any type of function approximation method can be applied in our research model. This includes point-to-point approximation models, global approximation, Hermitian interpolation approximation, and others. A robust approximation method holds the potential to significantly enhance model performance.

Here, we present only one approach to approximate time series data using Fourier series, as an example, and evaluate its performance. Readers are encouraged to adapt and explore different approximation methods to construct new models according to their own needs.

We focus primarily on two sequences from ADS-B: velocity $[ v_0, v_1, \dots v_{k-1}, v_k ]$, and altitude $[ h_0, h_1, \dots h_{k-1}, h_k ]$ on the time series $[ t_0, t_1, \dots t_{k-1}, t_k ]$. Assume that $k$ is big enough so that this series has enough points to calculate.

Normalize the data $t_{i}^* = {(t_i-t_0)}\pi/{T_M}$, mapping the time series $[ t_0^*, t_1^*, \dots t_{k-1}^*, t_k^* ]$ to the interval $[0, \pi]$, $T_M$ is a large constant.

Using linear interpolation to fit the function $h(t^*)$ in the interval $[0, \pi]$ as follows:
\begin{equation}
    \label{equ:fourier:h*}
    h(t^*) = 
    \begin{cases} 
        \frac{h_{i+1}-h_i}{t_{i+1}^*-t_i^*}\left( t^* - t_0^* \right) + h_i, & \text{if } t^*\in [t_i^*, t_{i+1}^* ),\quad i = 0,1,\dots ,k \\
        0, & \text{if } t^*\in [t_k^*, \pi ].
\end{cases}
\end{equation}

Map the function $h(t^*)$ to the symmetric interval $[-\pi, \pi]$, making it an even function on the symmetric interval $[-\pi, \pi]$, $h(t^*) = h(-t^*)$, if $t^*\in [-\pi, 0)$.

The Fourier series to approximate of $h(t^*)$ reads
\begin{equation}
    h(t^*) \approx \frac{\alpha_0}{2} + \sum_{n = 1}^{N_h}\left[ \alpha_n \cos{(nt^*)} + \tilde{\alpha}_n \sin{(nt^*)}   \right],\quad \text{if } t^*\in [-\pi, 0).
\end{equation}
Here $N_h$ is the number of terms in the Fourier series. Noting $h(t^*)$ is even on $[-\pi, \pi]$, we have $\tilde{\alpha}_n = 0$, and $a_n$ 
\begin{equation}
    \alpha_n = \frac{2}{\pi}\int_0^{\pi}h(t^*)\cos{(nt^*)}dt^*, \quad n = 0, 1,2,\dots,
\end{equation}
respectively. By the Riemann Lemma, $\displaystyle{\lim_{n\rightarrow +\infty} \alpha_n = 0}$.

Define
\begin{equation}
    \label{equ:fourier:Mi}
    \left\{
    \begin{aligned}
        &M(f)_j = \frac{f_{j+1}-f_j}{t_{j+1}^*-t_j^*},\\
        &Ct_j(n) = \cos{(nt_{j+1}^*)} - \cos{(nt_{j}^*)},\\
        &N(f)_j(n) = f_{j+1} \sin{(nt_{j+1}^*)} - f_{i} \sin{(nt_{j}^*)},
    \end{aligned}
    \right.
\end{equation}
$a_n$ reads
\begin{equation}
    \label{equ:fourier:an}
    \alpha_n = \frac{2}{\pi} \sum_{j=0}^{k-1}\left( \frac{M(h)_j*Ct_j(n)}{n^2}+\frac{N(h)_j(n)}{n} \right),
\end{equation}
where $n = 1,2,\dots$, and
\begin{equation}
    \label{equ:fourier:a0}
    \alpha_0 = \frac{2}{\pi}\sum_{j=0}^{k-1}\frac{(h_{j+1}+h_{j})(t_{j+1}^*-t_{j}^*)}{2}.
\end{equation}

Similarly, we can easily obtain the coefficients $\beta_n$ for fitting the ADS-B velocity sequence using the same method. Now, we have a good Fourier series approximation for $h(t)$. Here, as long as $N_h$ and $N_v$ are large enough, the Fourier series will have a small enough error with $h(t)$ and $v(t)$, respectively. That is, all the information in $h(t)$ and $v(t)$ can be fully expressed by a sequence $\alpha_n$ and $\beta_n$.

In addition to using Fourier series for approximation, readers can also choose methods such as Chebyshev polynomials, Z-transformed P\'{a}de rational approximation, and transforming the original sequence into a continuous sequence followed by Laplace transformation for Laurent series expansion. All of these approaches can effectively approximate ADS-B time sequences. These methods hold considerable potential for developing high-precision interval fuel consumption models.

\subsection{Interval Fuel Consumption Model}

Based on the above analysis, the interval fuel consumption $Q$ can be simplified into a multivariate real-valued function with $d+1$ parameters. There are various methods to approximate such a function with parameters. Here, we adopt the idea from~\cite{cheng2016wide}, where the parameters are used as the wide part, and the variables representing the nonlinear aspects of flight attitude are approximated using a deep model. Thus,
\begin{equation}
    \label{equ:interval:Q1}
    Q = \sigma\left( F_{\text{wide}}(t_0,\Xi) + F_{\text{deep}}(\mathbf{A}_{N_h},\mathbf{B}_{N_v})+bias \right).
\end{equation}
Here, $\sigma(.)$ is the sigmoid function, $F_{\text{wide}}(t_0,\Xi)$ computes the wide part, $F_{\text{deep}}(\mathbf{A}_{N_h},\mathbf{B}_{N_v})$ computes the nonlinear part, and $bias$ reads the the bias term, respectively.

The wide component is a generalized linear model of the following
\begin{equation}
    \label{equ:interval:Qwide}
    F_{\text{wide}}(t_0,\Xi) = \mathbf{w}_{\text{wide}}(t_0,\Xi)^T+b_{\text{wide}},
\end{equation}
$(t_0,\Xi)$ is a vector of $d+1$ features, $\mathbf{w}_{\text{wide}} = (w^{\text{wide}}_0,w^{\text{wide}}_1,\dots,w^{\text{wide}}_d)$  are the model parameters and $b_{\text{wide}}$ is the bias. The deep component is a feed-forward neural network, for computing the nonlinear part. Each hidden layer computes:
\begin{equation}
    \label{equ:interval:Qdeep}
    u^{(l+1)} = f(W^{(l)}u^{(l+1)}+b^{(l)}).
\end{equation}
Here $l$ mean the $l$-th layer, and $u^{(l)}$ is the node value in the $l$-th layer. The function $f(.)$ is the activation function. We can select $f(.)$ as a ReLU function. $W^{(l)}$ is model weights of the $l$-th layer and $b^{(l)}$ is the bias.

\subsection{Monotonic Interpolation and Instantaneous Fuel Consumption Model}

Considering both the monotonicity and the twice differentiability of the fuel quantity, the former ensures the positivity of instantaneous fuel consumption, while the latter guarantees the existence and boundedness of the derivative. 

Using the aforementioned algorithm, for interval fuel consumption, we obtain a time series concerning the fuel quantity $(T_0,Q_0,T_1,Q_1,\dots,T_K,Q_K)$ with monotonic condition $T_j<T_k,\quad Q_j<Q_k, \quad \forall 0\leq j<k\leq K$. We aim to obtain a smooth and monotonic function $Q(T)$.

Since the aforementioned time series are non-uniform, define $ I_j = \left[T_j,T_{j+1} \right]$, and the length of $I_j$ reads $\Delta T_j = T_{j+1}-T_{j}$. The interval derivative
\begin{equation}
    P_j=\frac{Q_{j+1}-Q_{j}}{\Delta T_j},\quad j = 0,1,\dots,K-1.
\end{equation}

Considering that the derivative of $Q(T)$ at the interval endpoints should not be approximated using the aforementioned interval derivatives, here we present a second-order approximation.
\begin{equation}
    \label{equ:insfuel:dQj}
    Q_j' = \frac{\Delta T_{j-1}P_j+\Delta T_jP_{j-1}}{\Delta T_{j-1}+\Delta T_j}, \quad Q_j'' = \frac{2P_j-2P_{j-1}}{\Delta T_{j-1}+\Delta T_j}.
\end{equation}
Here $\Delta T:=\max_j{\Delta T_j}$. The approximation error of $Q_j''$ is $O(\Delta T)$, and the error of $Q_j'$ is $O(\Delta T^2)$, respectively.

Discuss the interpolation approximation problem on $I_j$. Boundary conditions reads
\begin{equation}
    \label{equ:insfuel:Qduandian}
    \left\{
    \begin{aligned}
        & Q(T_j) = Q_j,\quad  Q(T_{j+1}) = Q_{j+1},\\
        & \frac{d}{dt}Q(T_j) = Q_j',\quad \frac{d}{dt}Q(T_{j+1}) = Q_{j+1}',\\
        & \frac{d^2}{dt^2}Q(T_j) = Q_j'',\quad \frac{d^2}{dt^2}Q(T_{j+1}) = Q_{j+1}'',
    \end{aligned}
    \right.
\end{equation}

Following the ideas of~\cite{cibulka2015three,fritsch1980monotone}, if we directly use polynomial functions for interpolation approximation on interval $I_j$, we will find it challenging to satisfy so many boundary conditions. To meet boundary conditions Eq.(\ref{equ:insfuel:Qduandian}), we divide $I_j$ into three sub-intervals.
\begin{equation*}
    I_{j1} = \left[T_j,T_{j+\frac{1}{3}} \right], \quad I_{j2} = \left[T_{j+\frac{1}{3}},T_{j+{\frac{2}{3}}} \right],\quad I_{j3} = \left[T_{j+\frac{2}{3}},T_{j+1} \right].
\end{equation*}
Here $T_{j+\frac{1}{3}}:=T_j+\frac{1}{3}\Delta T_j$, and $T_{j+\frac{2}{3}}:=T_j+\frac{2}{3}\Delta T_j$.

Define $Q_{j+\frac{1}{3}} = Q(T_{j+\frac{1}{3}})$, and $Q_{j+\frac{2}{3}} = Q(T_{j+\frac{2}{3}})$, and we calculate $Q(T)$ in $I_j$ reads
\begin{equation}
    \label{equ:insfuel:QTfinal}
    Q(T) = 
    \begin{pmatrix}
        Q^1(u) \\
        Q^2(v) \\
        Q^3(w)
    \end{pmatrix}
    = 
    \begin{pmatrix}
        Q_j & \frac{\Delta T_j}{3}Q_j' & Q_{j+\frac{1}{3}} & \frac{\Delta T_j}{3}b_j \\
        Q_{j+\frac{1}{3}} & \frac{\Delta T_j}{3}b_j & Q_{j+\frac{2}{3}} & \frac{\Delta T_j}{3}b_j \\
        Q_{j+\frac{2}{3}} & \frac{\Delta T_j}{3}c_j & Q_{j+1} & \frac{\Delta T_j}{3}Q_{j+1}'
    \end{pmatrix}
    \begin{pmatrix}
        H_0(.) \\
        H_1(.) \\
        G_0(.) \\
        G_1(.)
    \end{pmatrix},
\end{equation}
where $Q^1(u)$ is defined on $I_{j1}$, $u = 3(T-T_j)/\Delta t$, $Q^2(v)$ is defined on $I_{j2}$, $v = 3(T-T_{j+\frac{1}{3}})/\Delta t$, and $Q^3(w)$ is defined on $I_{j3}$, $w = 3(T-T_{j+\frac{2}{3}})/\Delta t$, respectively. The parameter can be calculated by
\begin{equation}
    \label{equ:insfuel:canshu}
    \left\{
    \begin{aligned}
        & a_j = Q_j'+\frac{\Delta T_j}{6}Q_{j}'',\quad d_j = Q_{j+1}'-\frac{\Delta T_j}{6}Q_{j+1}'',\\
        & 2b_j = a_j+\omega_j,\quad 2c_j = d_j+\omega_j,\\
        & \omega_j = 3P_j-(Q_j+Q_{j+1})+\frac{\Delta T_j}{9}(Q_{j+1}''-Q_{j}'') \\
        & Q_{j+\frac{1}{3}} = Q_{j}+\frac{\Delta T_j}{9}\left(Q_{j}'+\frac{3}{2}a_j+\frac{1}{2}\omega_j \right),\\
        & Q_{j+\frac{2}{3}} = Q_{j+1}-\frac{\Delta T_j}{9}\left(Q_{j+1}'+\frac{3}{2}d_j+\frac{1}{2}\omega_j \right).
    \end{aligned}
    \right.
\end{equation}
The Hermite polynomials $H_0$, $H_1$, $G_0$, and $G_1$ read
\begin{equation}
    \label{equ:insfuel:Hermite}
    \left\{
    \begin{aligned}
        & H_(t) = (1-t)^2(1+2t),\quad H_1(t)=t^2(3-2t),\\
        & G_0(t)=t(1-t)^2,\quad G_1(t) = -t^2(1-t).
    \end{aligned}
    \right.
\end{equation}
By leveraging the properties of Hermite polynomials, it is straightforward to verify that Eq.(\ref{equ:insfuel:QTfinal}) satisfies Boundary Conditions (\ref{equ:insfuel:Qduandian}), while also ensuring monotonicity and twice differentiability on $I_j$. Finally, by differentiating $Q(T)$, we can obtain the expression for instantaneous fuel consumption at any given point.

\section{Error Analysis }\label{sec:erroranalysis}

The dataset significantly impacts the performance of the model. In this section, we first introduce the dataset and the training methods used for the model. Then, we perform error analysis and convergence analysis on the given model using this dataset.

\subsection{Data Set and Model Training}

The data were collected from publicly available sources, including FlightRadar24, Umetrip, VariFlight, and some ACARS data.The data cover the period from April 2022 to December 2023, focusing on passenger flights departing from and arriving in mainland China. It can be summarized into three main parts which can be found in Table \ref{tab:dataset:summary}.

\begin{table}
\centering
\label{tab:dataset:summary}
\caption{123}
\begin{tabular}{l p{8cm}}
\hline\hline
\textbf{Data Part} & \textbf{Description} \\ \hline
Flight parameter: Flight Dynamic Data & Flight date, flight number, departure and arrival locations, departure and arrival times, registration mark number, aircraft type, wingspan, etc. \\ 
Trajectories: ADS-B & Flight date, flight number, registration mark number, timestamp, message type, longitude, latitude, altitude, speed, etc. \\ 
Fuel Data: ACARS & Flight date, flight number, registration mark number, timestamp, message type, remaining fuel in the tanks, etc. \\ \hline
\end{tabular}
\end{table}

It is important to note that, considering the difficulty of obtaining data, we only selected three parameters for the inherent flight characteristics: aircraft type, aircraft age, and wingspan. We acknowledge that other parameters, such as the flight's payload, also affect fuel consumption but were not included in this model due to data acquisition challenges. However, as we will see in the error analysis, even with just these three parameters and the ADS-B data, the calculation error remains within acceptable engineering limits. In this chapter, we focus on testing the model's performance without delving too deeply into the impact of other parameters.

For training data selection, we used ADS-B data from scheduled passenger flights between April 2022 and November 2023, with validation data from December 2023. Given that some aircraft might not have clean and usable ACARS data, we selected only 11 aircraft types for training, as shown in Table \ref{tab:dataset:aircraft_types}. Notably, these 11 aircraft types cover $90.88\%$ of the actual flights in Chinese airspace. Our dataset consists of 6.147 million flights.

\begin{table}
\centering
\caption{Aircraft Types Dataset}
\label{tab:dataset:aircraft_types}
\begin{tabular}{lrrr}
\hline\hline
\textbf{Aircraft Type} & \textbf{Number of Samples} & \textbf{Percentage (\%)} \\ \hline
Airbus 320/200         & 2,816,138                & 37.86              \\
Airbus 330/200         & 108,638                 & 1.46               \\
Airbus 319/100         & 413,928                 & 5.56               \\
Airbus 321/200         & 676,331                 & 9.09               \\
Airbus 330/300         & 99,304                  & 1.34               \\
Airbus 350/900         & 70,543                  & 0.95               \\
Boeing 737/800         & 2,078,163                & 27.94              \\
Boeing 737 MAX 8       & 115,465                 & 1.55               \\
Boeing 737/700         & 247,531                 & 3.33               \\
Boeing 777/300        & 52,481                  & 0.71               \\
Boeing 787/900        & 19,905                  & 0.27               \\
\hline
\textbf{Total}         & 74,381,235               & 100.00             \\ \hline
\end{tabular}
\end{table}

To enhance the accuracy of interval fuel consumption calculations, we also included segments of ADS-B training data, provided that ACARS data is available, to ensure more comprehensive model training and significantly increase the training data volume. A set of comparative experiments are conducted with the same experimental conditions to show the superior performance of our proposed model.

\subsection{Error Analysis}

We selected flight data from December 2023 as the test set to validate the model's performance. In this section, we primarily use relative error for the error analysis of the model. The relative error is expressed as follows:
\begin{equation}
    \label{equ:error:mape}
    \varepsilon = \frac{1}{N_{\text{test}}}\sum_{j = 1}^{N_{\text{test}}}\left| \frac{y^j_{\text{model}}-y^j_{\text{true}}}{y^j_{\text{true}}}\right|.
\end{equation}
Here, $y^j_{\text{model}}$ represents the value calculated by the model for the $j$-th test data in the test set, while $y^j_{\text{true}}$ represents the true value for the $j$-th test data in the test set. $N_{\text{test}}$ denotes the number of samples in this test.

\subsubsection{Convergence Analysis}

We analyzed errors and convergence in the Fourier Series Model by conducting numerical experiments with training datasets of varying sizes and Fourier truncation radii. Error analysis results are presented in Table \ref{tab:error_matrix},\ref{tab:error_matrix2} and Fig. \ref{fig:errorfourier}, where dataset sizes and error magnitudes follow an exponential growth pattern and are displayed on a logarithmic scale.

\begin{table}[h]
\centering
\caption{Error Matrix for Different Data Sizes (Part 1: $N = 10$ to $N = 50$).}
\label{tab:error_matrix}
\begin{tabular}{l*{6}{c}}
\hline\hline
\textbf{Data Size} & \textbf{$N=10$} & \textbf{$N=20$} & \textbf{$N=30$} & \textbf{$N=40$} & \textbf{$N=50$} \\
\hline
$21,629$       & 0.060540 & 0.059120 & 0.054860 & 0.055682 & 0.057476 \\
$43,623$       & 0.056205 & 0.053365 & 0.053291 & 0.052394 & 0.053963 \\
$81,016$       & 0.053066 & 0.051721 & 0.048806 & 0.048582 & 0.048955 \\
$174,614$      & 0.050899 & 0.048955 & 0.046190 & 0.045891 & 0.045966 \\
$340,248$      & 0.046638 & 0.044994 & 0.043499 & 0.042752 & 0.042901 \\
$646,249$      & 0.045592 & 0.042827 & 0.042004 & 0.041182 & 0.041108 \\
$1,315,670$    & 0.043574 & 0.041556 & 0.040510 & 0.039389 & 0.039389 \\
$2,700,088$    & 0.041481 & 0.039912 & 0.039090 & 0.038566 & 0.038342 \\
$5,610,242$    & 0.040136 & 0.037669 & 0.037445 & 0.036847 & 0.036249 \\
$11,228,756$   & 0.037584 & 0.035591 & 0.035093 & 0.034666 & 0.034523 \\
\hline
\textbf{Rate} & 0.073167 & 0.077314 & 0.070210 & 0.073265 & 0.079571 \\
\hline
\end{tabular}
\end{table}

\begin{table}[h]
\centering
\caption{Error Matrix for Different Data Sizes (Part 2: $N = 60$ to $N = 100$).}
\label{tab:error_matrix2}
\begin{tabular}{l*{5}{c}}
\hline\hline
\textbf{Data Size} & \textbf{$N=60$} & \textbf{$N=70$} & \textbf{$N=80$} & \textbf{$N=90$} & \textbf{$N=100$} \\
\hline
$21,629$       & 0.056504 & 0.056878 & 0.057999 & 0.058896 & 0.060017 \\
$43,623$       & 0.052767 & 0.053963 & 0.053216 & 0.055010 & 0.053664 \\
$81,016$       & 0.050525 & 0.050600 & 0.050376 & 0.051048 & 0.050749 \\
$174,614$      & 0.047610 & 0.046788 & 0.046638 & 0.048208 & 0.048432 \\
$340,248$      & 0.042229 & 0.042229 & 0.042303 & 0.042154 & 0.042378 \\
$646,249$      & 0.040883 & 0.041257 & 0.039912 & 0.040211 & 0.040360 \\
$1,315,670$    & 0.039389 & 0.039239 & 0.038641 & 0.038342 & 0.038342 \\
$2,700,088$    & 0.038193 & 0.037968 & 0.037894 & 0.037669 & 0.037296 \\
$5,610,242$    & 0.035801 & 0.035726 & 0.035726 & 0.035502 & 0.034829 \\
$11,228,756$   & 0.033740 & 0.033669 & 0.033455 & 0.033527 & 0.033100 \\
\hline
\textbf{Rate} & 0.081807 & 0.083877 & 0.085576 & 0.090846 & 0.093309 \\
\hline
\end{tabular}
\end{table}

\begin{figure}
    \centering
    \includegraphics[width=0.8\linewidth]{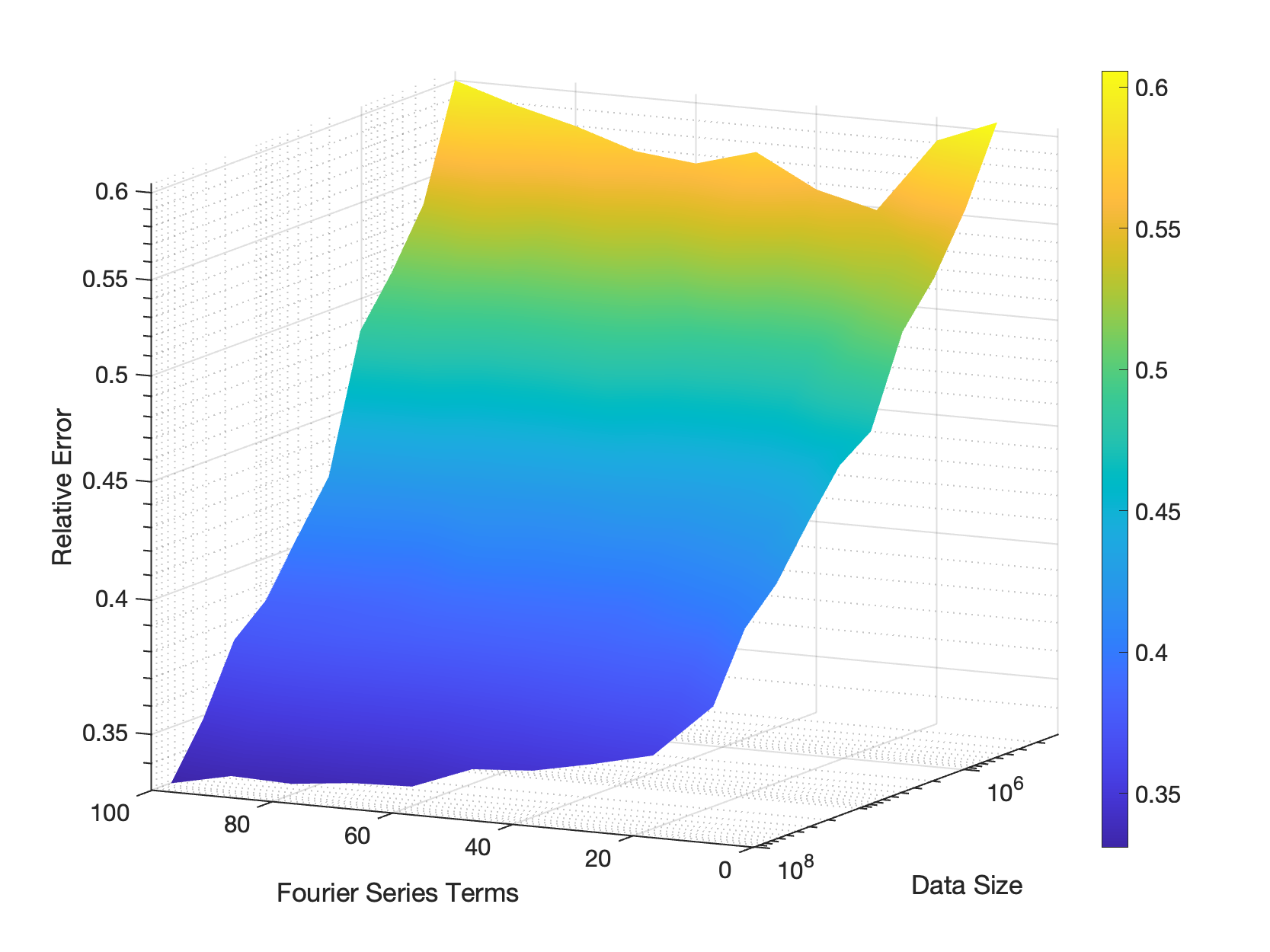}
    \caption{Errors distribution across different dataset sizes and parameter settings. Both the error axis and the axis for training data volume are plotted using a logarithmic scale. It can be observed that in logarithmic coordinates, the error exhibits a linear relationship relative to the training data volume. As the data volume increases, this linear relationship becomes more pronounced.}
    \label{fig:errorfourier}
\end{figure}

In our experiments, there is currently no reason to differentiate between $N_h$ and $N_v$. Therefore, we chose the truncation radius $N = N_h = N_v$. From Fig. \ref{fig:errorfourier} and Table \ref{tab:error_matrix},\ref{tab:error_matrix2} we can observe that, for a fixed training dataset, the overall error decreases as the truncation radius increases. However, there may be some fluctuations in smaller datasets. This phenomenon is likely due to the fact that larger truncation radii require more deep learning parameters, and the sparse data in smaller datasets do not sufficiently train models with larger truncation radii, leading to underfitting. When the training dataset size exceeds 1 million, models with larger truncation radii can still be well-trained, and the error continues to decrease.

Conversely, for a fixed Fourier series truncation radius, the error generally decreases as the training data size increases. This is evident in Fig.1, where for each fixed truncation radius model, the error decreases linearly with the increase in data size on a logarithmic scale. Even when the training data size reaches the order of 11 million, the error continues to decline.

To measure the rate of error decrease, we examine the convergence rate of the error with respect to the training data size, denoted the relative error $\varepsilon \sim O(N_{size}^m)$. Here, $m$ is the convergence rate and $N_{size}$ means the number of training data. By taking the logarithm of both sides of the equation and using the least squares method, we calculated the results shown in the last row of Table \ref{tab:error_matrix}. It can be seen that for smaller truncation radii, the convergence rate is around $0.07$. As the truncation radius increases, the convergence rate gradually increases, approaching $0.1$. This indicates that larger truncation radii can better fit the time series of altitude and velocity, thereby giving the model a greater chance of accurately calculating fuel consumption. This also indirectly proves the necessity of incorporating altitude and velocity time series for high-precision fuel consumption calculations.

From the current training data of $11$ million samples, if we choose a truncation radius of $100$, the model's performance can achieve an average error of $3.31\%$. This dataset approximately represents one and a half years of flight data, including ACARS and ADS-B data. If the dataset size were to double, we can expect the relative error to be $3.31\%/2^{1.0933} = 3.103\%$.

\subsubsection{Comparative Error Analysis Across Different Dimensions}

We assess model errors by flight duration and aircraft type, highlighting the model's strengths. For interval fuel consumption, we examine errors across different durations, finding that longer intervals facilitate more thorough flight posture analysis and better Fourier series approximations, reducing errors.

\begin{figure}
    \centering
    \includegraphics[width=0.8\linewidth]{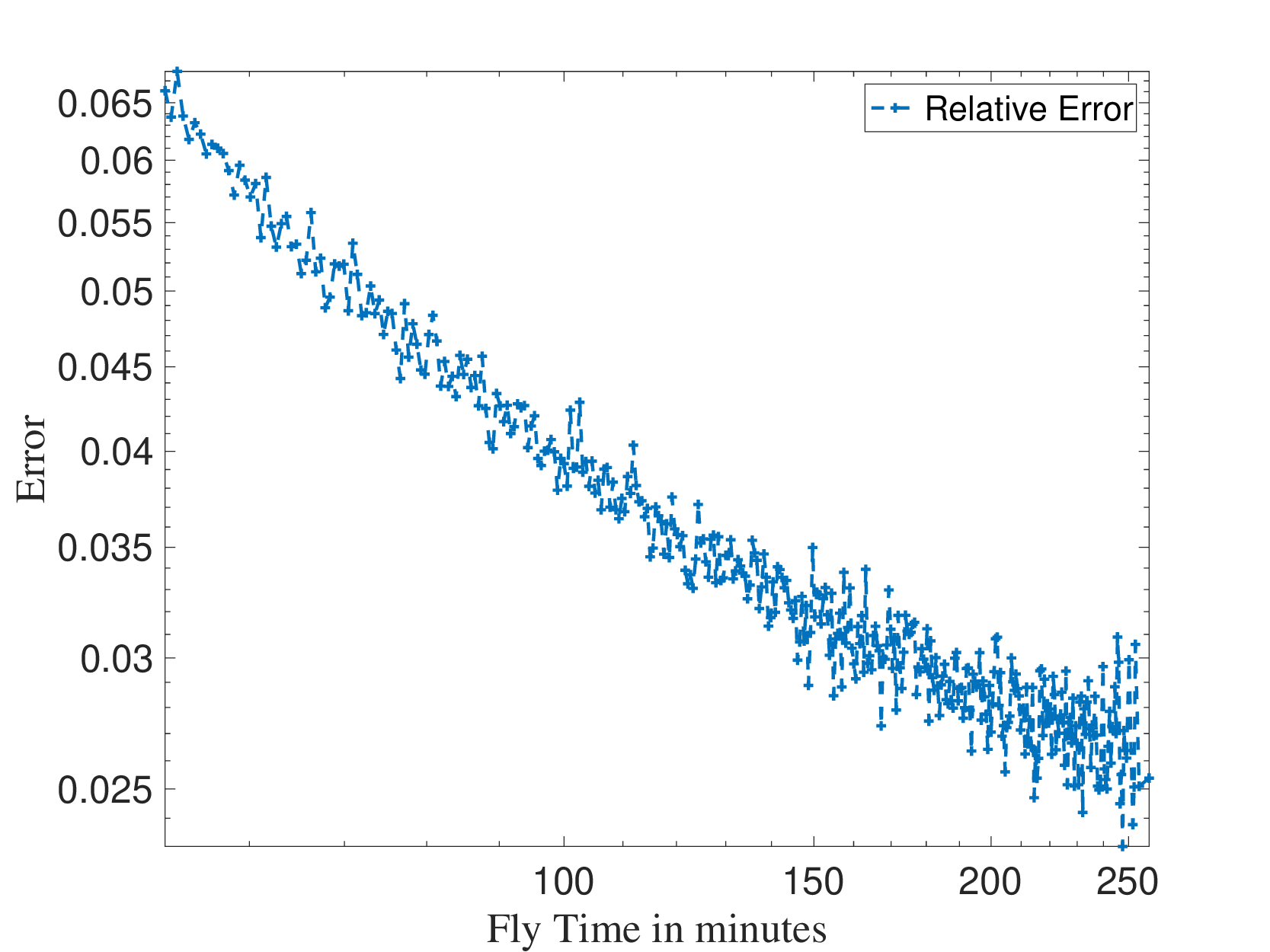}
    \caption{Error distribution across different flight duration. Both the horizontal and vertical axes are plotted on a logarithmic scale. It is evident that in logarithmic coordinates, the error decreases linearly relative to flight duration.}
    \label{fig:errorpoint}
\end{figure}

We will depict the errors in a scatter plot in Fig. \ref{fig:errorpoint}. The horizontal axis represents the flight duration, and the vertical axis represents the relative error. Both axes are shown in logarithmic scale. It can be observed that longer flight durations result in lower calculation errors, aligning closely with our hypothesis. Furthermore, on logarithmic coordinates, the error curve forms approximately a straight line with a slope of approximately $-0.5765$. This indicates that the error diminishes with increasing flight duration at a rate of $T^{-0.58}$.
We illustrate the distribution of errors across aircraft models in Fig. \ref{fig:erroraircraft}, showing that errors are generally under 4\% for most models. The Airbus A330-300 has errors over 5\%, but it comprises only 1.45\% of the fleet. Despite extensive data coverage over the year, the specific reasons for the A330-300's higher errors remain unclear.

\begin{figure}
    \centering
    \includegraphics[width=0.8\linewidth]{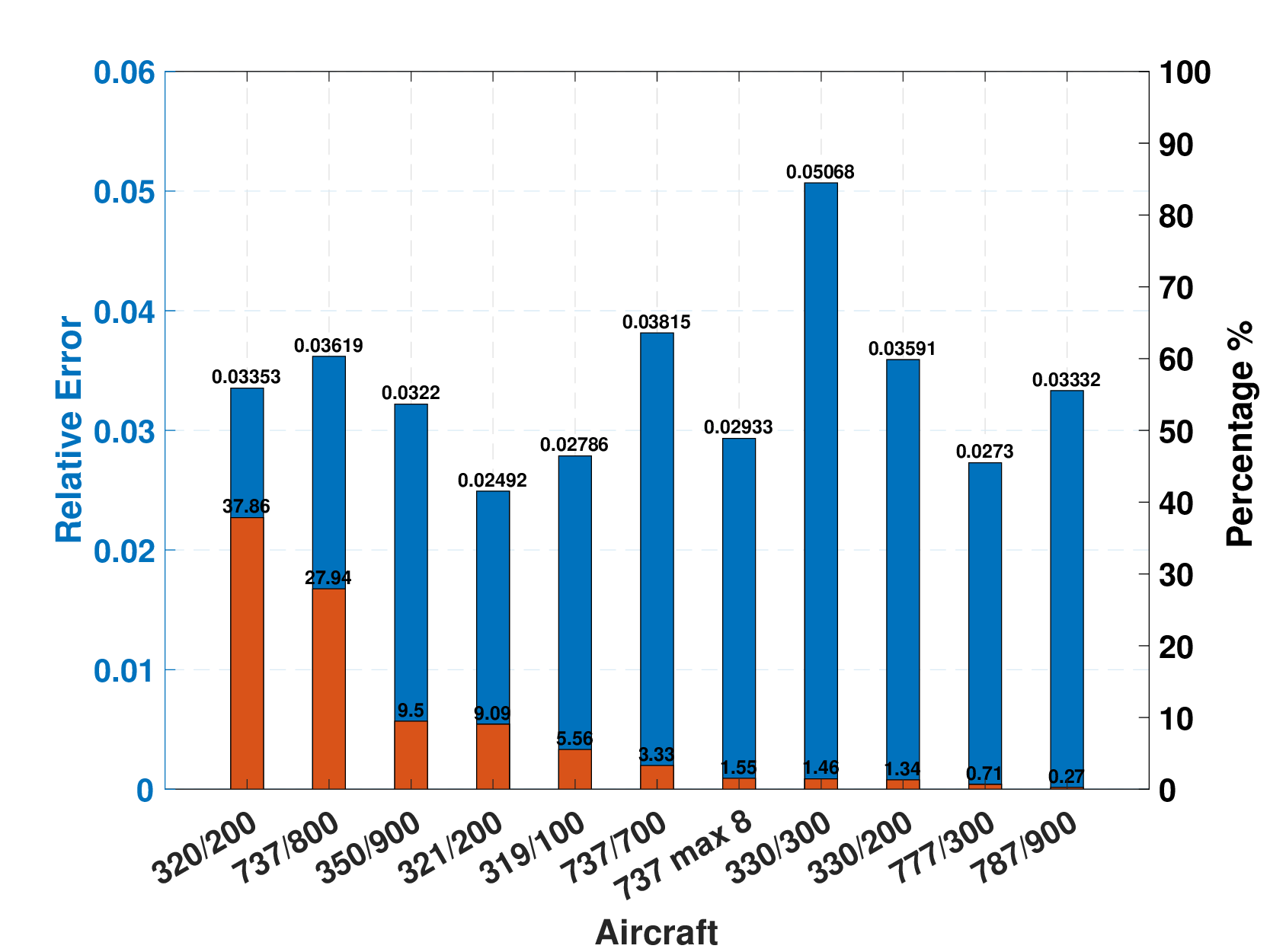}
    \caption{Error distribution across different aircraft types. The aircraft types shown here remain the top 11 mainstream models discussed earlier. Some general aviation and rare aircraft types are excluded from the scope of this study due to their limited data availability.}
    \label{fig:erroraircraft}
\end{figure}

After reorganizing and reviewing external data, we referenced Civil Aviation Leisure Station \footnote{Please refer to \url{http://xmyzl.com/}} for some insights. Among these 11 mainstream aircraft models in China, more than half are equipped with the same engine type. The Airbus A330-300, however, is notable for being equipped with multiple engine variants. It's confirmed that the A330-300 is a twin-engine aircraft. Different versions are powered by three different engines: the Pratt \& Whitney PW4000, Rolls-Royce Trent 700, and CF6-80E1. Our training and testing data are concentrated between 2022 and 2023. Among the aircraft operating within China during this period, we encountered only the PW4000 series and RR Trent 700 engines. The PW4000 sub-type we encountered was the PW4170, which differs in thrust configuration from the Rolls-Royce Trent 772B-60. Fuel consumption varies accordingly with flight posture, which we did not explicitly include in our data. It's anticipated that introducing engine models as a separate variable could significantly improve the accuracy of our calculations in addressing these error issues.

\subsubsection{Comparative Error Analysis of Different Models}

In this part, we compare mainstream data-driven models and computing frameworks used in constructing our model. Actual fuel consumption data continues to be validated against ACARS data. The validation dataset consists of randomly sampled data from the full year of 2023. Errors are outputted in terms of MAPE (Mean Absolute Percentage Error), as shown specifically in Table. \ref{tab:error_compare}. Here, C denotes corrections made using actual ADS-B flight mileage.

\begin{table}[h]
\centering
\caption{Summary of MAPE error for different Models. Here C means Actual Flight Distance Correction. EEA Model can refer to \cite{EEA2023}, ICAO Model can refer to \cite{ICAO2024} and ES\&T 2022 Model can refer to \cite{zhang2022developing}, respectively. To validate the importance of deep mining of speed and altitude sequences for accurately predicting flight fuel consumption, we further developed a machine learning model based on the LGB model and trained it using the same training data.}
\label{tab:error_compare}
\begin{tabular}{l c}
\hline\hline
\textbf{Model} & \textbf{MAPE Error} \\
\hline
EEA Model without 'C' & 0.232397 \\
EEA Model with 'C' & 0.197122 \\
ICAO Model without 'C' & 0.142689 \\
ICAO Model with 'C' & 0.177321 \\
ES\&T 2022 Model with 'C' & 0.162587 \\
LGB Model without Flight Attitude & 0.121419 \\
Fourier with $N = 50$ & 0.034523 \\
\hline
\end{tabular}
\end{table}

The EEA model has errors over 20\% on flight datasets, reduced to 19.7\% after mileage correction. The ICAO model shows a 14.2\% error, increased by crude mileage correction, whereas \cite{zhang2022developing} achieve 13.2\%. Our model maintains a consistent error of about 3.4\%, indicating a significant improvement. We also developed a Light GBM-based model excluding flight posture data, which shows a 12.1\% error rate as shown in Table \ref{tab:error_compare}
}. This emphasizes the importance of considering flight posture to reduce computational errors.

We highlight our models' advantages in error analysis, derived from understanding the physical aspects of flights, improving fuel consumption calculations.

\subsubsection{Error Analysis of Instantaneous Fuel Consumption Models}

Finally, we evaluate the performance of the instantaneous fuel consumption model. Based on the analysis above, shorter intervals increase the error of the interval fuel consumption model but enhance the accuracy and effectiveness of interpolation. It is crucial to select an effective interval length for computing interval fuel consumption and utilize a monotonically smooth interpolation method for calculating instantaneous fuel consumption.

\begin{figure}
    \centering
    \includegraphics[width=0.9\linewidth]{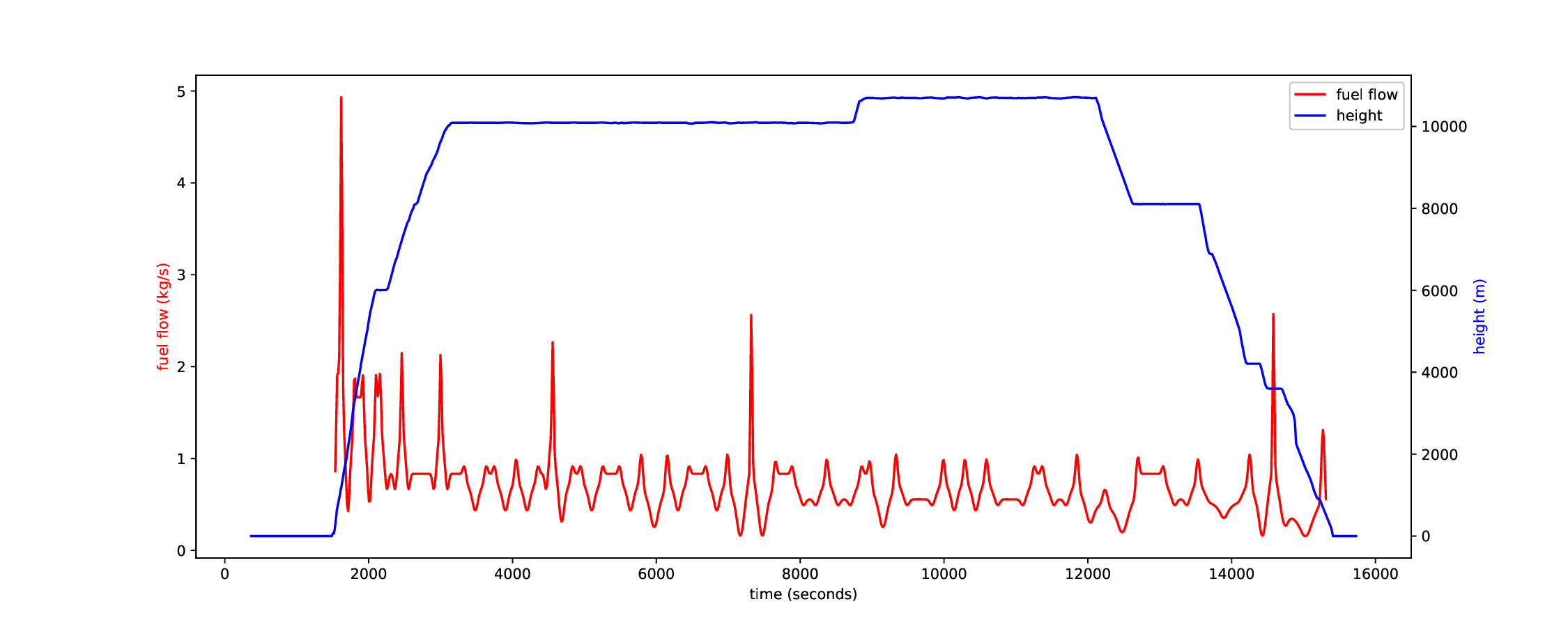}
    \caption{Calculation results of instantaneous fuel consumption for a certain flight.}
    \label{fig:fueliterpolate}
\end{figure}

Through multiple experiments, we have found that, on average, a 200-second interval is generally suitable for computing interval fuel consumption segments. The specific computational results are depicted in Fig. \ref{fig:fueliterpolate}. The red curve represents the instantaneous fuel consumption profile, while the blue curve denotes the altitude of the aircraft.

As a validation of effectiveness and comparison, we utilize a subset of QAR data as reference and calculate the relative L2-error defined as follows:

\begin{equation}
    \label{equ:error:l2error}
    \varepsilon_{L_2}[f(t)] = \frac{||f(t)-f_{\text{ref}}(t)||_2}{||f_{\text{ref}}(t)||_2} = \frac{\sqrt{\int_0^{T_0}{|f(t)-f_{\text{ref}}(t)|^2dt}}}{\sqrt{\int_0^{T_0}{|f_{\text{ref}}(t)|^2dt}}}.
\end{equation}

Here, $f(t)$ is the computed instantaneous fuel consumption curve, and $f_{\text{ref}}(t)$ is the reference instantaneous fuel consumption curve. Due to the significant difficulty in acquiring QAR data, we cannot present statistically meaningful results here, but only illustrate an example from a flight. For this particular flight, the calculation error $\varepsilon_{L_2} =0.088674$, slightly higher than the interval fuel consumption error.

\section{Application in Aviation Carbon Emissions}\label{sec:application}

Calculating carbon emissions from fuel consumption is the primary method to account for aviation carbon emissions. Our analysis includes a horizontal carbon emission map and vertical emission distribution, essential for studying their environmental impacts. The horizontal aspect involves mapping emissions across regions, as highlighted by \cite{li2022mapping}, who studied city-level emissions in 2019 but noted challenges due to missing precise location data. As detailed in \cite{vennam2017modeled}, understanding the vertical distribution of emissions is crucial, particularly in terms of how pollutants are transported between the lower stratosphere and upper troposphere. \cite{barrett2010global} also showed that atmospheric circulation causes emissions to travel varying distances, with emissions during landing, takeoff, and cruise phases shifting up to 900 kilometers due to these dynamics. However, coarse resolution cannot accurately study the transport effects of pollutants from aviation fuel.

\begin{figure}
    \centering
    \includegraphics[width=1.0\linewidth]{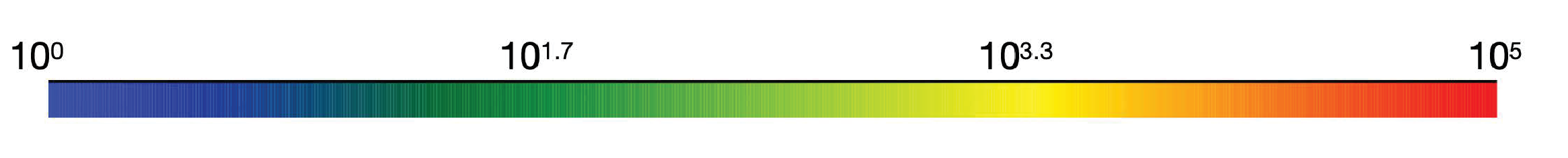}
    \caption{Color bar for aviation carbon emissions.}
    \label{fig:color}
\end{figure}

\begin{figure}
    \centering
    \includegraphics[width=0.3\linewidth]{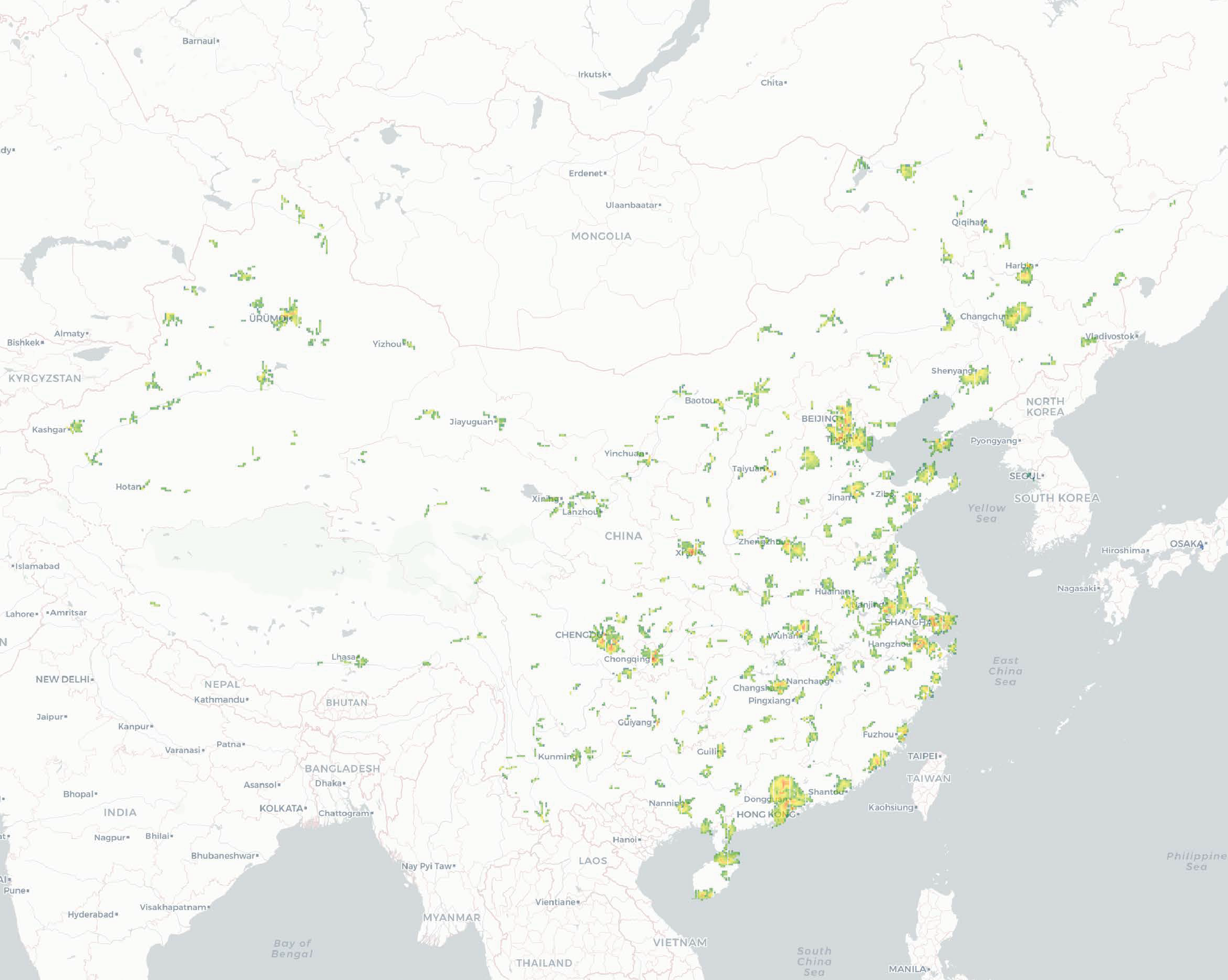}
    \includegraphics[width=0.3\linewidth]{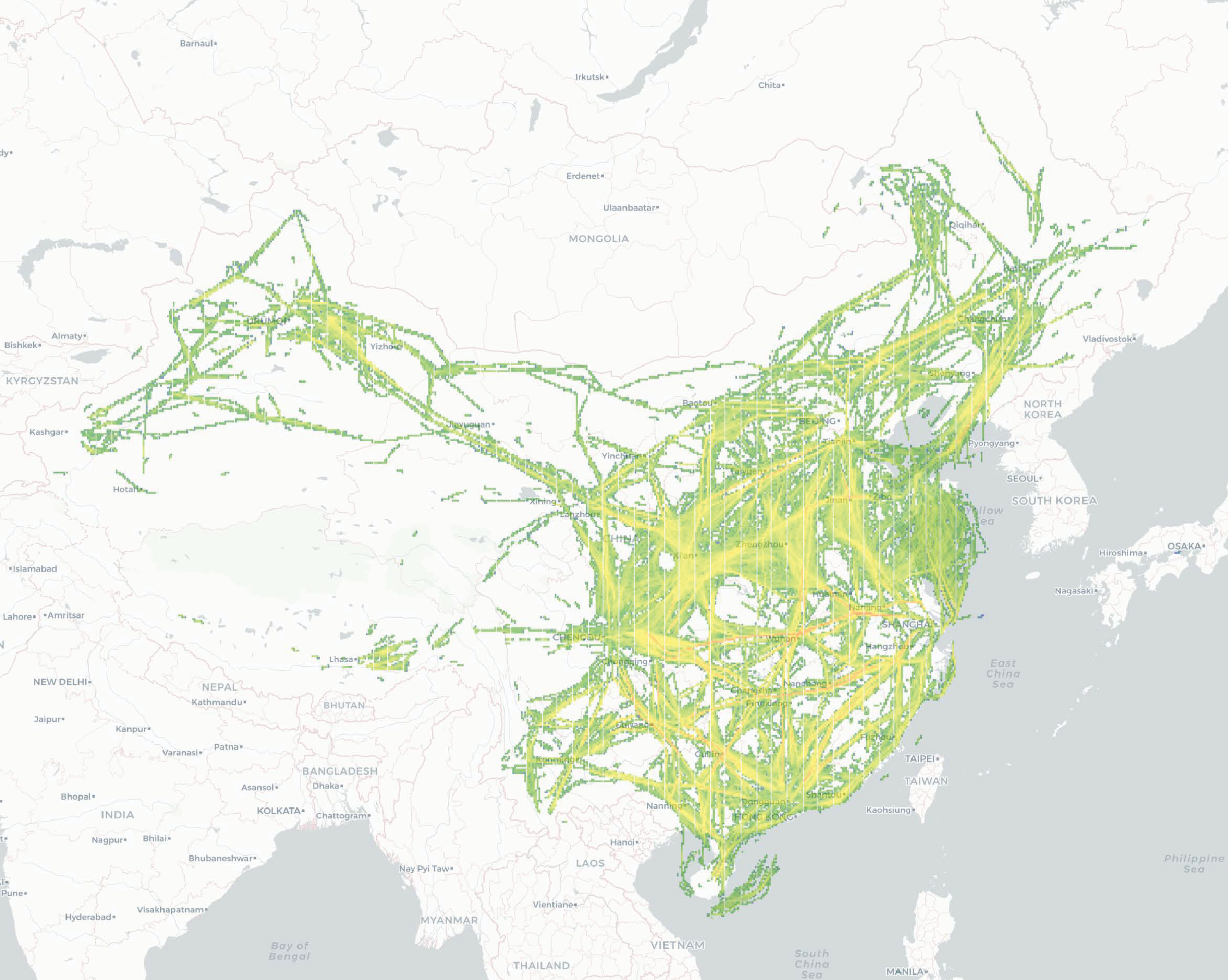}
    \includegraphics[width=0.3\linewidth]{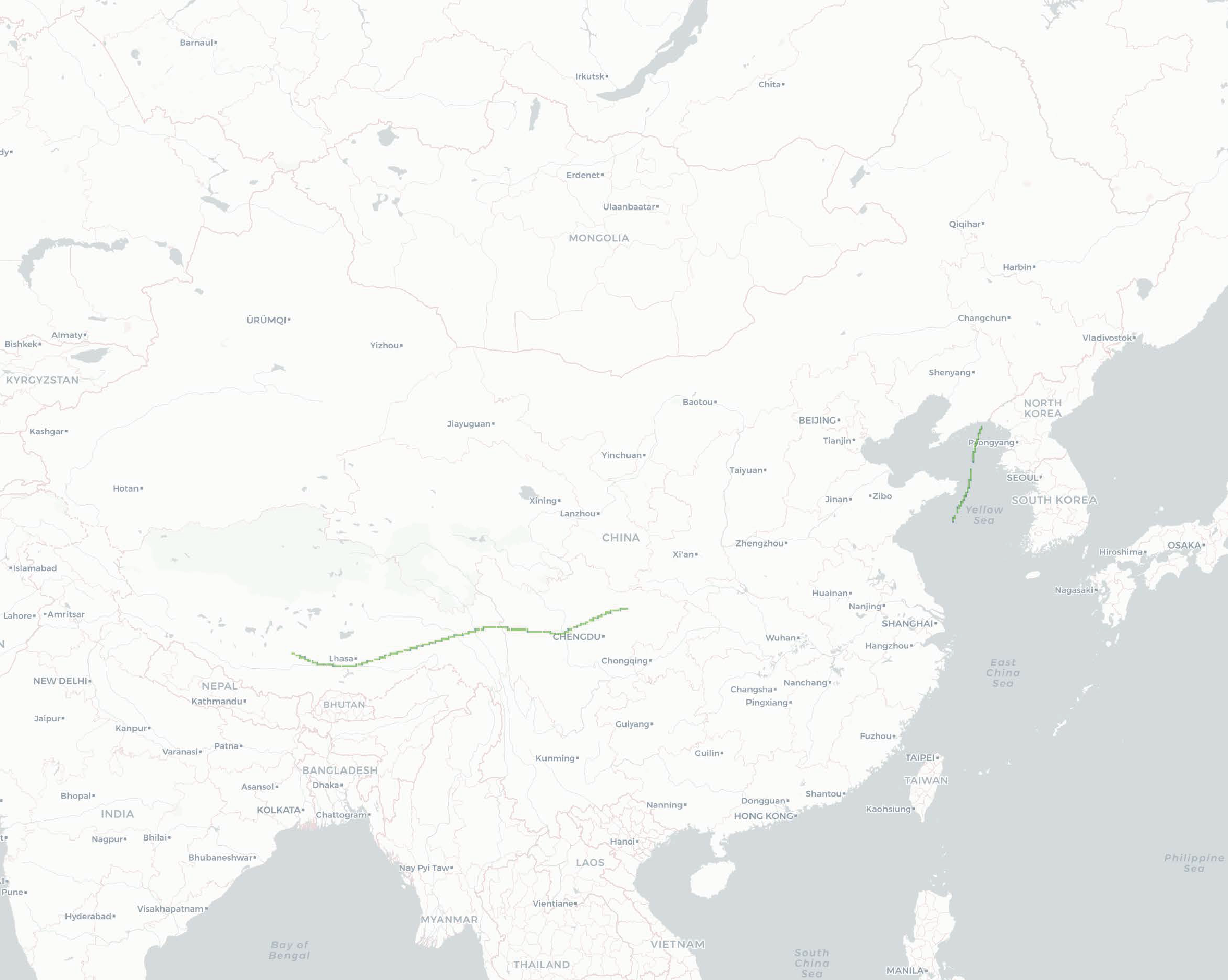}
    \caption{Spatial distribution of aviation carbon emissions across different altitude layers.}
    \label{fig:test1}
\end{figure}

Given the significant horizontal and vertical aspects of aviation emissions, our application combines flight trajectories and instantaneous fuel consumption to showcase a high-precision, high-resolution example of carbon emissions on July 7, 2024. These data enable the matching of spatial planes with altitude levels and the accounting of carbon emissions by provinces and cities. This provides robust support for further advancing spatial grid research and in-depth studies of carbon emissions transportation. We present maps of aviation carbon emissions at different flight altitudes (1000-2000 meters, 8000-9000 meters, and above 13000 meters) in Fig.\ref{fig:test1}, each corresponding to distinct flight phases and atmospheric conditions. These maps are enhanced with a color bar in Fig.\ref{fig:color}, providing a visual representation of emission intensity at each altitude layer. This aids in quickly assessing variations in carbon output across different flight stages.

This map is divided into altitude segments of 1,000 kilometers, with an emission resolution accuracy of 0.33 degrees per grid point. 

Here, we only provide an example of both horizontal and vertical aviation carbon emission distributions that can only be calculated using our algorithm. Additional altitude layer emission maps and analyses are available in the appendix.

\section{Conclusion}\label{sec:Conclusion}
We developed a comprehensive mathematical framework for aviation fuel consumption using ADS-B data to track flight attitudes like acceleration and descent, enhancing traditional calculations. We first transform the problem into a multivariate function fitting task, and utilizes function approximation techniques to progressively improve the approximation of the flight trajectory sequence. Then using a deep neural network (DNN) to capture the complex relationships between flight trajectory and fuel consumption. By training the DNN with comprehensive data, we obtain high-precision fuel consumption estimates. Additionally, we developed a second-order smooth monotonic interpolation method to ensure smoothness and realism in instantaneous fuel consumption estimates, enhancing the overall accuracy and reliability of the model. Numerical experiments were conducted to validate the effectiveness of the proposed model. The results demonstrate that our approach significantly reduces the average error of interval fuel consumption to as low as $3.31\%$. Furthermore, the error in the integral sense of instantaneous fuel consumption is reduced to $8.86\%$. These figures represent a substantial improvement over traditional methods, underscoring the accuracy and reliability of our model.

As a state-of-the-art solution, our model achieves the lowest estimation error for aircraft fuel consumption reported in current literature. This advancement has important implications for both theoretical research and practical applications. For instance, airlines can utilize our model to enhance their fuel efficiency strategies, optimize flight routes, and reduce operational costs. Moreover, the improved accuracy in fuel consumption estimation can support more precise environmental impact assessments and the formulation of effective carbon emission reduction policies.

\bibliographystyle{apalike}
\bibliography{fuel2024.bib}  

\end{document}